\documentclass[useAMS,usenatbib]{mn2e}

\usepackage{graphicx}
\usepackage{amsmath,amssymb}
\usepackage[usenames,dvipsnames]{color}
\usepackage{aas_macros,natbib}
 \usepackage[pdftex,
 				colorlinks=true,
 				citecolor=red,urlcolor=blue,
 				pdfpagelabels=true,
 				pdfstartview=Fit,
 				pdftitle={Can Planck constrain indirect detection of dark matter in our galaxy?},
 				pdfauthor={Delahaye, Boehm \& Silk},
				bookmarksopen=true,
				]{hyperref}

\newcommand{\beq}{\begin{equation}}
\newcommand{\eeq}{\end{equation}}
\newcommand{\bea}{\begin{eqnarray}}
\newcommand{\ena}{\end{eqnarray}}

\title{Can Planck constrain indirect detection of dark matter in our galaxy?}

\author[T. Delahaye~\textit{et al.}]{Timur Delahaye$^{1}$,C\'eline B\oe hm$^{2,3}$ and Joseph Silk$^{4,5}$\\
$^{1}$Instituto de F\'isica Te\'orica UAM/CSIC
Universidad Aut\'onoma de Madrid
Cantoblanco, 28049 Madrid, Spain\\
$^{2}$Inst. for Particle Physics Phenomenology, Durham University, South Road, DH1 3LE, United Kingdom\\
$^{3}$LAPTH, CNRS/UMR 5108, 9 chemin de Bellevue - BP 110, 74941 Annecy-Le-Vieux, France\\
$^{4}$Astrophysics department, Oxford University, Keble Road, OX1 3RH, United Kingdom\\
$^{5}$Institut d'Astrophysique de Paris, CNRS/UMR7095, 98 bis boulevard Arago, 75014 Paris, France
}

\begin{document}
\pagerange{1--5} \pubyear{2011}
\date{\today}
\maketitle

\begin{abstract}
We investigate the synchrotron emission (both intensity and morphology) associated with generic dark matter particles and make predictions for the PLANCK experiment 
using the FERMI data and a model for the astrophysical sources. Our results indicate that the morphology of the 
dark matter plus astrophysical source synchrotron emission is frequency-dependent. We show that a thorough comparison between LFI and HFI data can potentially provide a new tool for constraining 
the dark matter particle mass. 
%and therefore potentially compete with present dark matter direct detection and FERMI experiments. Hence, instead of using a template to subtract the galactic diffuse synchrotron emission in PLANCK data, 
%a safer approach requires to thoroughly compare the LFI and HFI ``synchroton'' maps. Indeed, small 
%differences could reveal evidence for dark matter.
\end{abstract}

\begin{keywords}
Astroparticle physics -- dark matter -- radio continuum: ISM.
\end{keywords}

\section{Introduction}
There are many indirect detection techniques that can be used to elucidate the nature of the dark matter in our Universe. Among them, the anomalous production of cosmic rays and $\gamma$-rays, first proposed by \cite{Silk:1984zy} in the context of self-annihilating neutralinos, has received much attention
%,Rudaz:1987ry,Kamionkowski:1990ty,Rudaz:1987ry,2008ApJ...682..400P,Silk:1985ax,Hagelin:1986gv,Turner:1986vr,Bergstrom:1988fp,Bouquet:1989sr,bens,Beacom:2004pe,Bringmann:2007nk} and $\gamma$-rays (\textit{e.g.} Ref.~\cite{Turner:1986vr,Bergstrom:1988fp,Bouquet:1989sr,bens,Beacom:2004pe,Bringmann:2007nk}) have been studied at 
%length since the 1980's and received even more attention
since the PAMELA experiment confirmed an excess of positrons at relatively low energies. 
A new type of dark matter signature has also been proposed in the form of 
% as they may turn out to be important. For example, looking for an
 anomalous radio emission from leptonic annihilation products
\citep{bens,Colafrancesco:2005ji,Zhang:2008rs,Borriello:2008gy,2009arXiv0911.1124B,Crocker:2010gy,2011MNRAS.410.2463S,Bergstrom:2008ag,2010arXiv1008.5175B}.
% In particular, observations towards the Galactic Centre 
%may constrain or exclude light annihilating dark matter particles in the 1-10 GeV range, depending on their annihilation cross section. 
%Seemingly, the observation of an anomalous $\gamma$-ray emission from dwarf Spheroidal galaxies has already proven to be efficient to constrain 1-1000 GeV dark matter particles 
%(still depending on the annihilation cross section) \cite{Abdo:2010ex} and could even constrain dark matter particles as light as a few MeV \cite{Hooper:2003sh}.
%Other signatures, \textit{e.g.} a dark matter induced Sunyaev-Zel'dovich (SZ) effect, could also be potentially interesting but they are somewhat more controversial \cite{Colafrancesco:2004sp,Yuan:2009yy,Lavalle:2009fu,Boehm:2008nj}. 
%
%While the latest results demonstrate that current experiments (\cite{Adriani:2008zr,Collaboration:2008aaa,Abdo:2009zk} )are becoming sensitive enough to test interesting dark matter 
%models \cite{Cirelli:2009dv,2010arXiv1012.0588Z,Abdo:2010ex,Ascasibar:2005rw,Ibarra:2009nw,Papucci:2009gd,Zhang:2009ut,2010arXiv1006.0477B}), 
% important developpments are yet to come. 
It may even be possible to discriminate decaying from annihilating dark matter scenarios using 
the morphology of the electromagnetic emission if a signal were detected \citep{Ascasibar:2005rw,2010PhRvL.105v1301B}. 
%Alternatively, with the use of forthcoming data, one may exclude a whole class of Dark Matter models. 
%
%Whatever the results from indirect detection experiments, they will have to be confronted 
%to the data which are being collected in underground experiments. 
%Several (\cite{2011PhRvL.106m1301A,Bernabei:2008yi,Bernabei:2010mq,2010Sci...327.1619C,2011arXiv1103.4070E}) 
%have recently announced detection of signals or events which might be related to the presence of 
%dark matter particles in our halo. If a signal was to be confirmed, one could certainly exploit the complementarity between these different techniques 
%to determine the dark matter mass and total pair annihilation cross sections. In addition, now that LHC is also running, one could eventually access the physics beyond the Standard Model.

There has been recent interest in the exploitation of galactic foregrounds in all-sky CMB experiments for
%One new method that has not been fully exploited yet though is 
the detection of synchrotron light (corresponding to microwave and submillimetre radiation) emitted by a new, relativistic, population of 
electrons originating from dark matter annihilations or decays.  It was pointed out a few years ago that the subtraction of known foregrounds (extrapolated from the Haslam data at 408 MHz and 
from Parkes at 2.4 GHz) to microwave frequencies showed a residual trace  
in the 22GHz channel of WMAP.  It was then speculated that the origin of the so-called WMAP haze \citep{Finkbeiner:2003im} could be due to dark matter particles 
\citep{2007PhRvD..76h3012H,Dobler:2007wv}, although more recent investigations on its nature now seem to favour an astrophysical interpretation rather 
than a dark matter origin \citep{2010ApJ...724.1044S}. 
Whatever the origin of the WMAP haze, this work has  demonstrated that dark matter could potentially be seen in CMB experiments via galactic foregrounds. 

Here, we demonstrate that exploiting PLANCK data may open up a new window for indirect searches of dark matter particles and 
offers a way to cross-check the results obtained from other channels. 
Our assumptions are the following: i) we suppose that dark matter particles annihilate or decay into electrons (and positrons) with some specified branching ratio, as 
%(this enables to perform a model independent approach as was done also 
in \cite{2010PhRvL.105v1301B} and \cite{2010arXiv1006.0477B}
ii) we use a semi-analytical approach to solve the diffusion equation (as described in \cite{Delahaye:2007fr}) to propagate relativistic electrons, 
and iii) we assume a smooth NFW dark matter halo profile, parameterized as:
$$
f_{DM}(\textbf{r}) =\frac{R_\odot}{\textbf{r}}  \left(\frac{R_\odot + R_\Gamma}{\textbf{r} + R_\Gamma}  \right)^2,
$$
with $R_\Gamma =$~20~kpc and $R_\odot =$~8.5~kpc. 
%Such a cut is justified by HFI and LFI angular resolutions which are about 5' for the highest frequencies and 33',24',14' for the 30,44,70 GHz frequency bands. 
%
%
%In Sec.\ref{section1}, we recall the calculation of the synchrotron light emission originating from both the dark matter and  astrophysical sources. 
%In Sec.\ref{section2}, we predict the morphology and intensity of this emission for candidates with a mass of 10, 40 and 200 and 800 GeV. We will consider, depending on the case,  
%values of the magnetic field of 3, 6 and 25 $\mu$G as well as the ''MED'' and ''MAX'' set of propagation parameters. We conclude in Sec.\ref{section3}.
%

%===========================================================================
%===========================================================================
\section{Method \label{section1}} 
%===========================================================================
%===========================================================================

In order to estimate the synchrotron emission from dark matter particles, we first need to determine the electron/positron production rate by dark matter ($Q_n$) for each adopted  particle mass $m_{dm}$. 
For this purpose, we  exploit the fact that the same population of relativistic electrons is expected to produce both  synchrotron radiation relevant for CMB experiments 
and a measurable cosmic ray flux at the Sun's position relevant for balloon or satellite experiments such as FERMI.

%In order to estimate the synchrotron emission from dark matter particles, we first need to determine the relevant part of the dark matter parameter space. 
%For this purpose, we can exploit the fact that the same population of relativistic electrons is expected to produce both a synchrotron light relevant for CMB experiments 
%and a signal in the FERMI experiment. The combination of both data actually set a constraint on the ratio $Q_n/m_{dm}$. 

%The important point is that, 
For a given dark matter mass, one cannot arbitrarily increase the value of $Q_n$ because this would lead to an excess of electrons and positrons in the FERMI data.
Before evaluating the ``dark'' synchrotron emission, we thus need to determine the maximal allowed dark matter pair annihilation 
cross-section which is compatible with the FERMI data for specific values of the dark matter mass \footnote{Note that maximising the annihilation cross-section (or the decay rate) may actually lead to some structures in the cosmic ray flux but this is not in conflict 
with observations. Besides, local sources may induce similar effects.}.  

This, however, requires  knowledge of  the background emission.~{\it I.e.}~one has to estimate the electron and positron flux  expected from galactic astrophysical sources. 
Unfortunately there is no exhaustive catalogue of high energy electron emitters in the galaxy. 
Hence, a technique, used in particular by WMAP to remove the 
synchrotron emission and access the cosmic microwave background (CMB),  assumes that radio (low frequency synchrotron) maps (\textit{e.g.} the Haslam maps at 408 MHz) 
are also valid at higher frequencies, rescaling them by a factor which is determined by requiring that the galactic synchrotron contribution obtained at microwave/sub-millimeter frequencies 
is consistent with the CMB data already in hand. In practice, this is equivalent to  assuming that i) any source which would contribute to the synchrotron emission at 
radio frequencies also gives a signal at higher frequencies and ii) the morphology of the emission stays the same whatever the frequency.

Here, we will show that these assumptions are not necessarily valid. Very high energy electrons, as produced by dark matter annihilations or decays can -- if energetic enough -- 
contribute to the highest frequencies only (\textit{i.e.} not produce any radio signal). Besides, as already demonstrated in \cite{2010JCAP...10..019M}, taking into account spatial and energy propagation 
can change the morphology of the electron distribution in the galaxy as well as the dependence of the electron flux with the energy and lead, in particular, to the wrong interpretation of the WMAP haze.

Given these drawbacks, we will adopt a different approach. Instead of extrapolating the Haslam maps 
from the lowest to the highest frequencies, we will assume that high energy electrons are produced by a smooth distribution of steady astrophysical sources 
and work out the flux that is expected at the Earth's position and at the FERMI energies by including  spatial and energy propagation.  
We will then compute the cosmic ray flux due to dark matter annihilations (or decays) and adjust the production rate so that the sum of both components does not exceed the electron flux measured by FERMI.
%We will then use the difference between the measured electron $+$ positron flux observed by FERMI and our preditions of the electron $+$ positron flux from the dark matter 
%to constrain its pair annihilation cross section at a given mass.

Among the most recent source distributions in the literature, the most statistically 
significant can be found in \cite{2004IAUS..218..105L} and can be described as follows:
\begin{equation} 
f_{\mathrm{L04}}(r,z) = \left(\frac{r}{R_\odot}\right)^{2.35} \exp\left(-\frac{r}{R_\odot}\right) \exp\left(-\frac{\vert z\vert}{0.1~\textrm{kpc}}\right).
\nonumber
\end{equation}
Fermi acceleration theory as well as radio observations of supernova remnants indicate that the spectrum of the 
electrons emitted by astrophysical sources is a power-law with a high energy 
exponential cut-off. Therefore, we will parameterize the source term as:
\begin{equation}
Q_{\rm SNR}(\epsilon,r,z) = A f_{\mathrm{L04}}(r,z) \epsilon^{-\sigma} e^{-\frac{\epsilon}{2\mathrm{TeV}}} \displaystyle
\end{equation}
where $\epsilon$ is the energy of the electrons,  $A$ is the amplitude and $\sigma$ the spectral index (expected to be slightly greater than 2). 
In the following, we will fix the parameters $A$ and $\sigma$ so as to give the best qualitative fit to the FERMI data with astrophysical sources only \citep{2009PhRvL.102r1101A}.

The propagation of high energy electrons in the galaxy can be well modelled in the two-diffusion-zone model (see for example \cite{1998ApJ...493..694M,2001ApJ...555..585M} or \cite{2008PhRvD..77f3527D}, 
for a complete description of the propagation model used here) with three parameters ($L$ the diffusion slab half-thickness, $\delta$ and $K_0$ the diffusion parameters). 
If one considers pulsars as a viable source of cosmic rays (see for instance \cite{2010A&A...524A..51D}), 
all the data which has been collected for local fluxes of electrons and positrons can be described by a given set of parameters. Owing to the energy losses, these data correspond 
to electrons and positrons emitted mostly in the nearby 2~kpc. 

Since the value of the magnetic field $B$ (in particular) is expected to vary significantly in the diffusion zone, 
one cannot guarantee that this set of parameters is valid throughout the whole Galaxy. However, since it is difficult 
to describe accurately such  variations and embed them into a semi-analytical approach, we will assume that the field is constant over the entire diffusion zone and consider the same 
set of propagation parameters whatever the source of high energy electrons. Finally, we neglect the reacceleration of cosmic rays while this may actually be important for low mass dark matter particles.

We can now estimate the synchrotron emission from dark matter particles in a smooth dark halo. 
The surface brightness
% \timur{I am not sure a surface brightness can be expressed in a specific 
%direction in$sr^{-1}$ for a diffuse emission, I would call this only the brightness.}
 is given by: 
$$I_{\nu}(l,b) = \Psi_{(l,b)} \times \frac{E}{\nu} \times b_{\rm sync.}(E) $$
with 
\begin{eqnarray}
  \Psi_{(l,b)}  &=&  \frac{N_e\ Q_n}{{\eta_n} \ b(E)}  \ \int d s(l,b) \int d^3x \  \left(\frac{\rho(x)}{m_{dm}}\right)^n \nonumber \\
&&\times G(\odot,x \leftarrow \epsilon,E_{inj}) \label{ne}
\end{eqnarray}
the electron flux at an energy $E$ and in a given direction (integrated along the line of sight). The term $G(\odot,x \leftarrow \epsilon,E_{inj})$ represents the Green's function. 
It encodes the propagation of the electrons (spatial diffusion and losses) from their place of ``birth'' to a position $x$ and from an injection energy $E_{inj}=n \times m_{dm}/2$ to a 
lower energy $E_{min}=\epsilon$. $N_e$ is a multiplicity factor.  Since dark matter always produces both an electron and a positron simultaneously, $N_e=2$. 
The term $b(E)$ accounts for the energy losses (inverse Compton and synchrotron) and $b_{\rm sync.}(E)$ for those due to synchrotron only. 
%Interestingly enough, when the magnetic field is large enough,  the losses equate the power radiated by the synchrotron 
%emission, and we find that: $$b(E) \simeq P(E) = \frac{1}{6 \pi} \ \sigma_T \ c \ \beta^2 \ \gamma^2 \ B^2.$$ 

The convention displayed in Eq.\ref{ne}, \textit{i.e.} $n=1,2$, denotes the decaying and annihilating DM cases respectively. 
The term $Q_1$ is the decay rate (expressed in $s$), $Q_2$ is the annihilation cross section (expressed in $\rm{cm^3/s}$), $\rho$ is the dark matter mass distribution and $m_{dm}$ is 
the dark matter mass. The term $\eta_n$ is equal to unity when dark matter is decaying ($\eta_{n=1} = 1$). It is equal to 2 or 4 if dark matter is annihilating ($\eta_{n=2}= 2$ if 
dark matter is a Majorana and  $\eta_{n=2}= 4$ if dark matter is a Dirac particle). 

Following this procedure, we find that for 40 GeV particles, the annihilation cross section can be as large as $\sigma v \simeq 1.5-2.5 \ 10^{-26} \rm{cm^3/s}$ in our galaxy without being in conflict with the 
FERMI data. This suggests that  annihilations in the primordial Universe were either occuring mostly into  particles other than electrons (and positrons) or 
the velocity-dependent term in the pair annihilation cross section into electrons is important  ($\sigma v = a + b v^2$ with $a>b$).   
For 100 GeV particles, the annihilation cross section is about $\sigma v  \simeq 7 \ 10^{-26} \rm{cm^3/s}$. This is somewhat larger than the canonical thermal annihilation  value required to explain all the dark matter today
(namely $3 \ 10^{-26} \rm{cm^3/s}$) but is still compatible with the FERMI measurement of the electron $+$ positron flux in the Milky Way. 
Such a $\sigma v$ value could suggest scenarios in which the 
annihilation cross section is enhanced in the galaxy due to the small velocity dispersion of the dark matter particles in the halo (c.f. the Sommerfeld enhancement). Hence constraints from spheroidal dwarf galaxies (dSph) may apply.

Although the FERMI limits on dark matter candidates obtained from dSph are stringent, they do depend on the dark matter mass and most notably on the adopted dark matter profile.
Using PLANCK data would therefore provide additional constraints and a means to cross check the FERMI results.

%============================================================================
%===========================================================================
\section{``Dark'' synchrotron emission \label{section2}} 
%===========================================================================
%===========================================================================

In what follows, we will display the most significant synchrotron map predictions. We focus on annihilating dark matter particles.
We use the ``MED'' (corresponding to $L=4$ kpc, $\delta=0.7$, $K_0= 0.0112$ $\rm{kpc}^2/\rm{Myr}$) 
and ``MAX'' (corresponding to $L=15$ kpc, $\delta=0.46$, $K_0= 0.0765$ $\rm{kpc}^2/\rm{Myr}$) set of propagation parameters. 
As demonstrated in our previous work \cite{2010PhRvL.105v1301B}, a smaller diffusion zone (corresponding to the ``MIN'' set of parameters) 
will lead to a more confined ``dark matter''synchrotron emission (brighter in the centre and fainter outside) while a more optimistic model of propagation (``MAX'') 
would lead to a brighter emission at larger latitude and longitude. 
Of course, the relative brightness of the emission at each frequency is affected by the choice of propagation parameters but, in this Letter, we do not attempt to perform a detailed analysis of 
the propagation parameters. We only point out that if propagation of cosmic rays in our galaxy is correctly described by the ``MED'' and ``MAX'' parameter sets, 
PLANCK may have the ability to constrain the dark matter mass.

To produce the dark matter-related synchrotron maps, we assume a monochromatic emission (\textit{i.e.}  one frequency corresponds to a single value of the electron energy). 
The relation between injection energy and  frequency then reads:
$$\nu_{max} = 16 \ \rm{MHz} \ \times \left(\frac{n}{2}\right)^2 \times \left(\frac{m_{dm}}{ \rm{GeV}}\right)^2 \times  \left(\frac{B}{\mu G}\right).$$ 
This well-known relation indicates that small dark matter masses cannot ``shine'' at high frequencies unless the magnetic field is very strong. 
Although obvious, this property turns out to be very important for dark matter searches.

\begin{figure}
	\centering
		\includegraphics[width=0.45\textwidth]{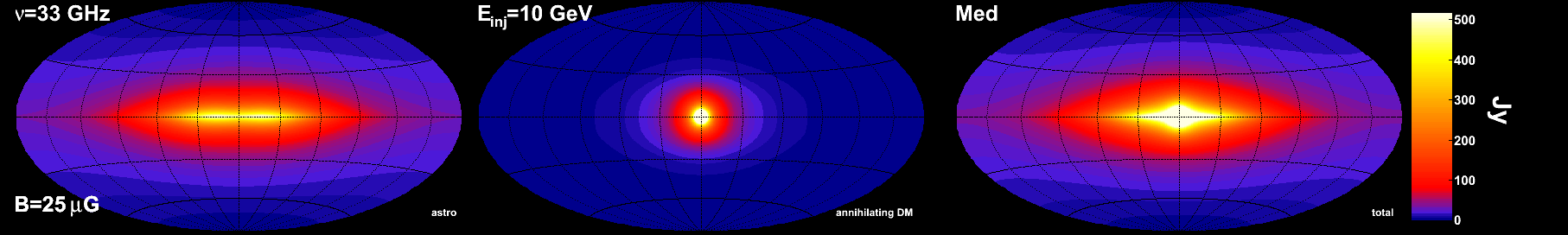}
	\caption{Synchrotron maps for 10 GeV dark matter particles, $B=25\mu$G. We use the MED parameter set and assume annihilating particles. 
	The emission from astrophysical sources is displayed in the left column; the dark matter prediction is shown in the middle panel and the sum of the two contributions is 
	dispayed in the right panel.}
	\label{fig:25muG_143GHz_Med_10GeV_ann}
\end{figure}

\begin{figure}
	\centering
	\includegraphics[width=0.45\textwidth]{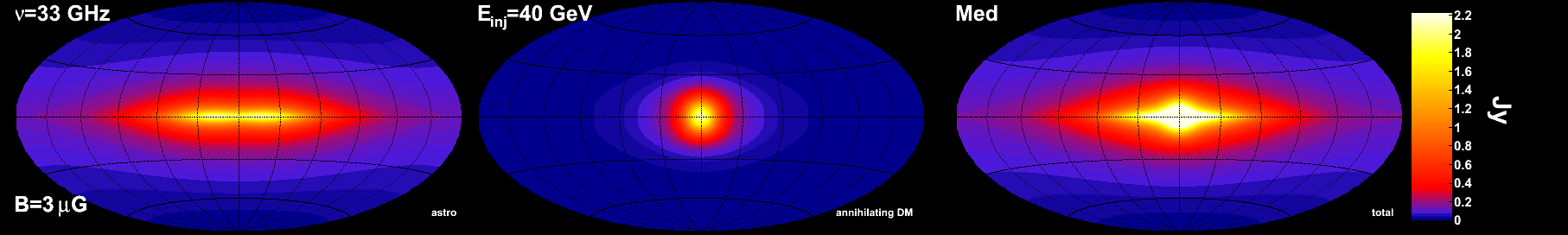}
		\includegraphics[width=0.45\textwidth]{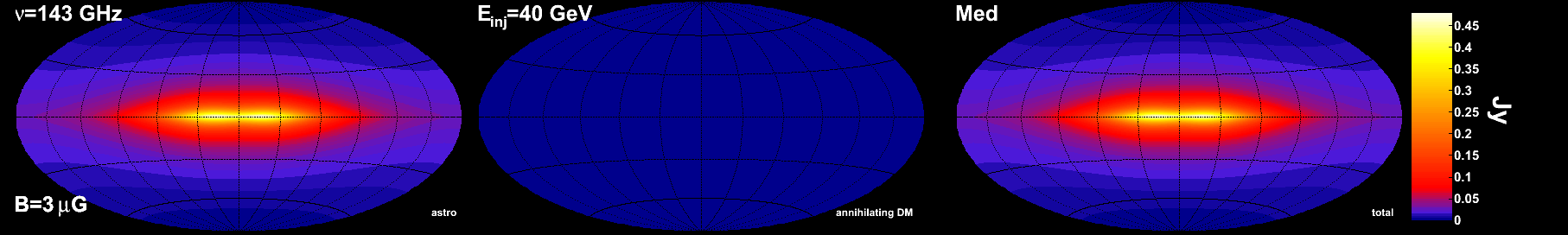}
		\includegraphics[width=0.45\textwidth]{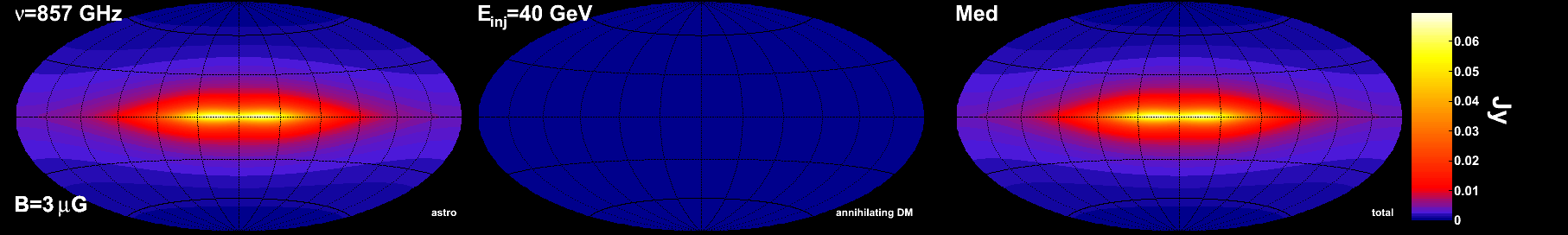}
	\caption{Synchrotron maps for 40 GeV dark matter particles, $B=3 \mu$G. We use the MED parameter set and assume annihilating particles.}
		\label{fig:3muG_143GHz_Med_40GeV_ann}
\end{figure}

In Fig.~\ref{fig:25muG_143GHz_Med_10GeV_ann}, we show that 10 GeV dark matter can shine at 33 GHz if the magnetic field is about 25 $\mu$G.
However, no signal is expected at higher frequencies unless the magnetic field is stronger. The intensity of the emission is large enough 
to be within the reach of PLANCK sensitiviy. The dark matter signal is very bright at the centre, as can be expected from the large value of the magnetic 
field (the latter indeed confines the electrons in the centre). However the sum of the two contributions is bright enough at high latitudes to have a chance of being detected by the LFI. 
This is consistent with previous dark matter analyses performed in the context of the WMAP haze \citep{Hooper:2010im}. Interestingly enough, for such parameters  
one also expects a radio signature in the galactic centre. As shown in \cite{bfs,2010arXiv1008.5175B}, one expects the radio emission to be about ten times smaller than 
the emission attributed to the central black hole. Therefore, in principle, the estimate of the radio emission should set a stronger limit on the cross-section. 
\textit{I.e.} it is likely to constrain cross-sections greater than $\sigma v \simeq 2 \ 10^{-27} \ \rm{cm^3/s}$. Nonetheless, one still expects a visible signal in PLANCK/LFI and no signal in HFI.

\begin{figure}
	\centering
		\includegraphics[width=0.45\textwidth]{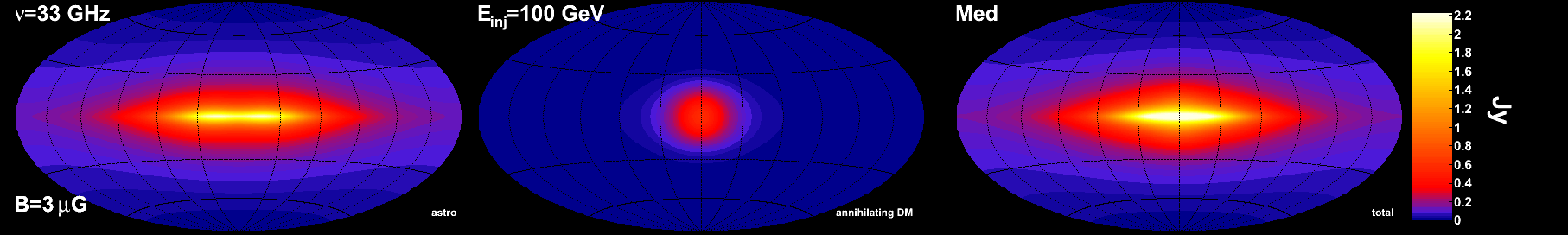}
		\includegraphics[width=0.45\textwidth]{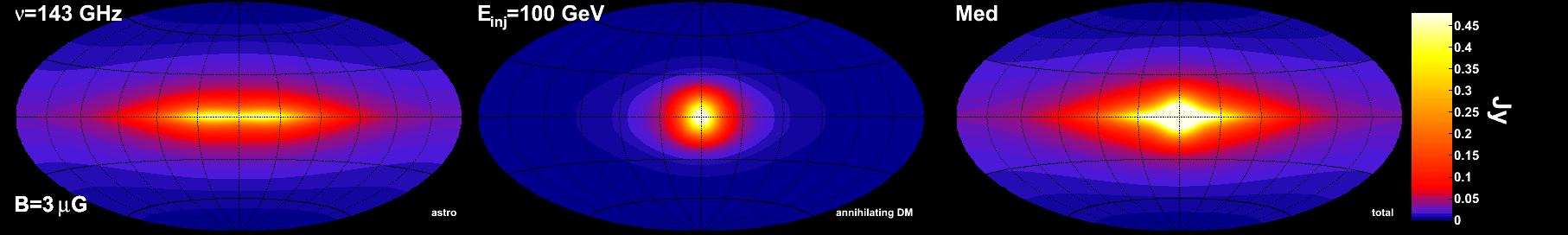}
		\includegraphics[width=0.45\textwidth]{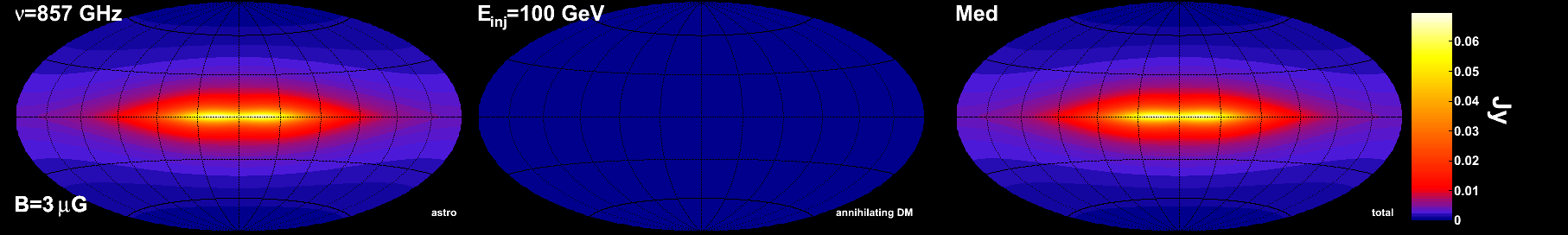}
	\caption{Synchrotron maps for 100 GeV dark matter particles, $B=3 \mu$G. We use the ``MED'' parameter set and assume annihilating particles. }
	\label{fig:3muG_143GHz_Med_100GeV_ann}
\end{figure}

\begin{figure}
	\centering
	\includegraphics[width=0.45\textwidth]{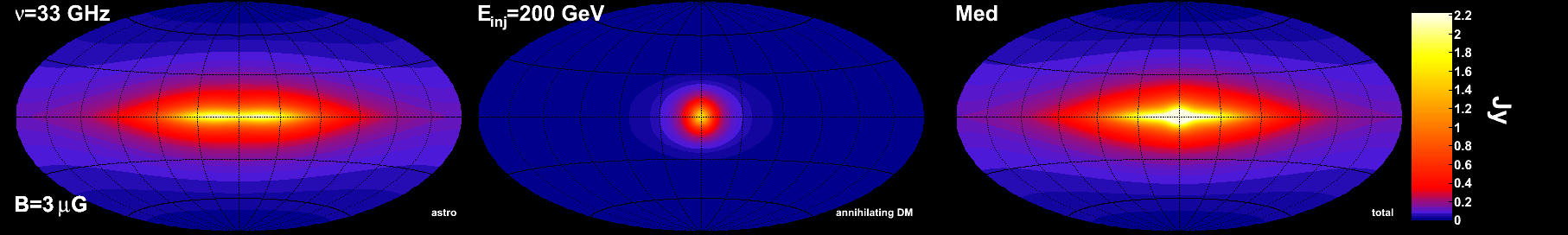}
		\includegraphics[width=0.45\textwidth]{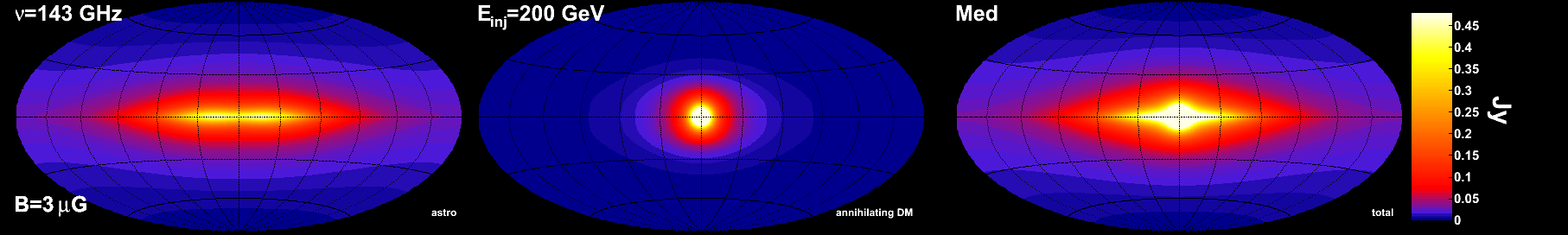}
		\includegraphics[width=0.45\textwidth]{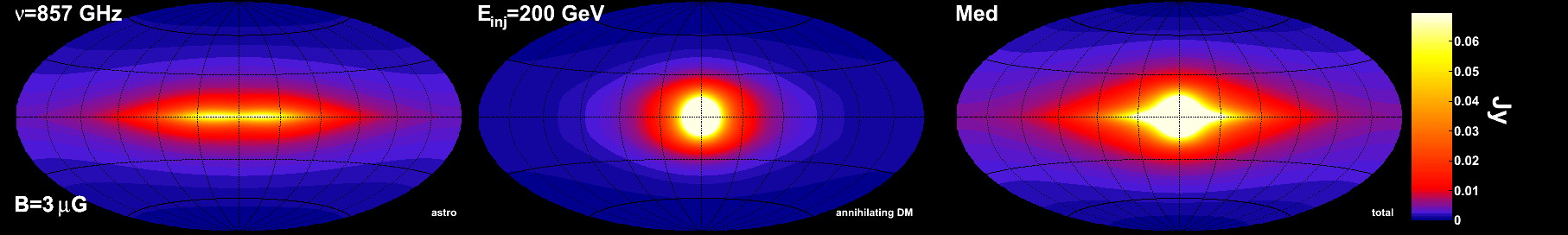}
	\caption{Synchrotron maps for 200 GeV dark matter particles, $B=3 \mu$G. We use the MED parameter set and assume annihilating particles.}
	\label{fig:3muG_143GHz_Med_200GeV_ann}
\end{figure}

When the mass is about 40 GeV and the magnetic field is close to the average value in the whole galaxy (cf. Fig.~\ref{fig:3muG_143GHz_Med_40GeV_ann}), 
one observes an extinction of the dark matter contribution to the synchrotron emission at large frequencies. 
This was to be expected from the frequency-energy relation but it does demonstrate again that comparing maps in different frequency channels is important. 
At 33 GHz, the sum of the astrophysical and dark matter contribution becomes visible close to the galactic centre  at high latitudes, and it should still be within the reach of LFI sensitivity. 
Finding the dark synchrotron contribution will be difficult but possible, and it is therefore important to compare all frequency channels 
before removing the radio maps extrapolated to high energies.

\begin{figure}
	\centering
	\includegraphics[width=0.45\textwidth]{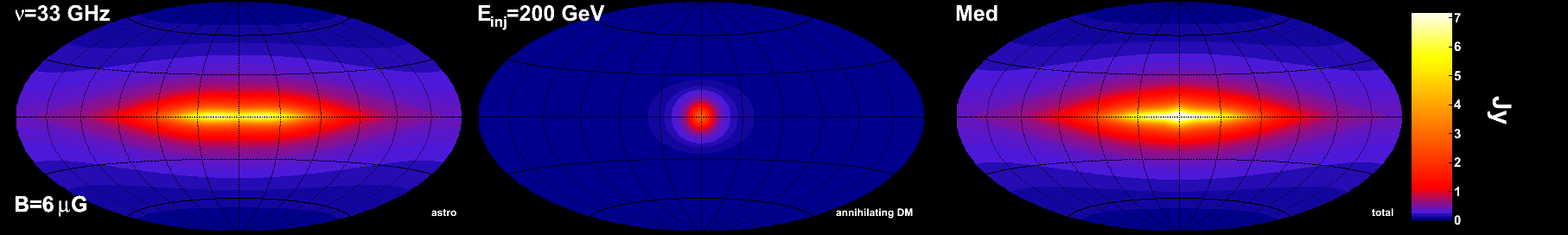}
		\includegraphics[width=0.45\textwidth]{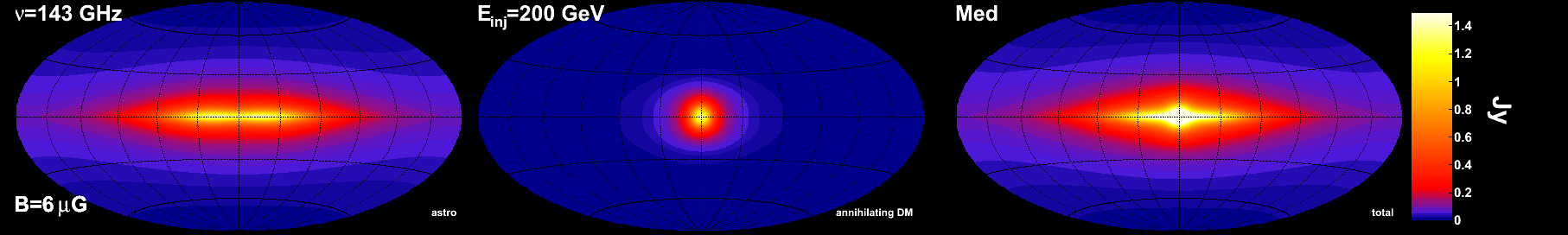}
		\includegraphics[width=0.45\textwidth]{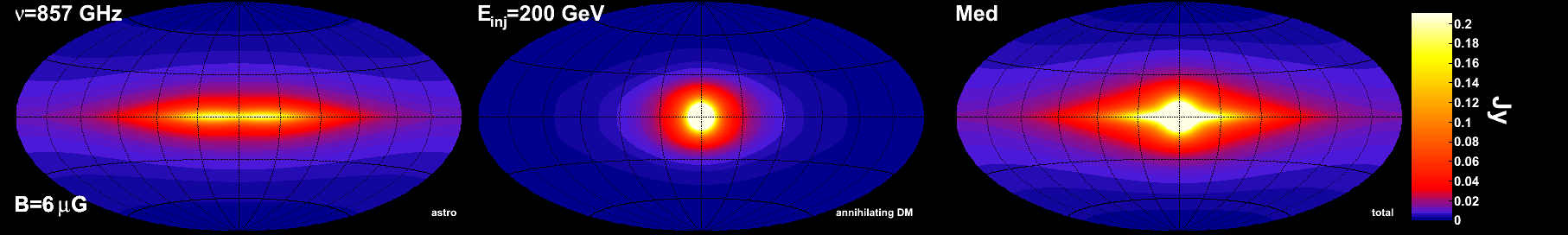}
	\caption{Synchrotron maps for 200 GeV dark matter particles, $B=6 \mu$G. We use the MED parameter set and assume annihilating particles.}
	\label{fig:6muG_143GHz_Med_200GeV_ann}
\end{figure}

The same features can be seen for 100 GeV (cf Fig.\ref{fig:3muG_143GHz_Med_100GeV_ann}),  except that the 33 GHz channel  actually seems less anomalous than the 143 GHz channel 
while there should be no visible signal at very large HFI frequencies. This illustrates how important it is to perform a thorough comparison of the synchrotron emission 
in the different frequency channels. Since the emission is expected to be about a few Jy, detecting the dark synchrotron emission would also be difficult but perhaps 
feasible and rewarding.

\begin{figure}
	\centering
		\includegraphics[width=0.45\textwidth]{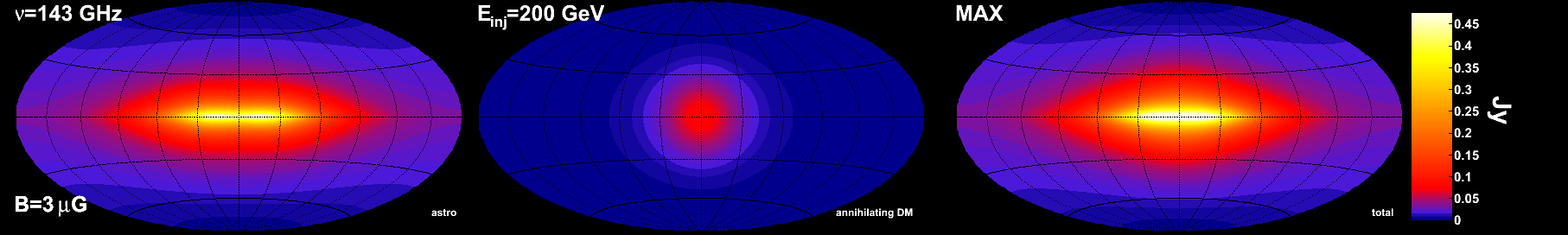}
		\includegraphics[width=0.45\textwidth]{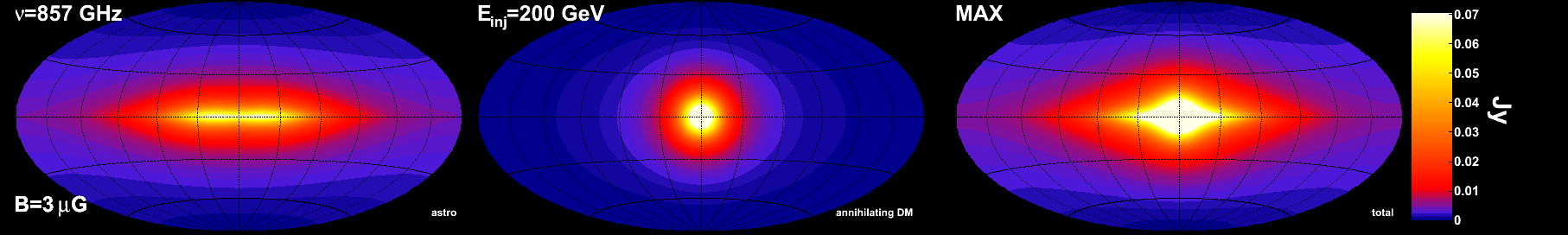}
	\caption{Synchrotron maps for 200 GeV dark matter particles, $B=3 \mu$G. We use the MAX parameter set and assume annihilating particles. }
	\label{fig:3muG_143GHz_MAX_200GeV_ann}
\end{figure}

At 200 GeV and $B=3 \mu$G (cf Fig.~\ref{fig:3muG_143GHz_Med_200GeV_ann}), we observe an interesting effect: namely extinction of the dark synchrotron emission at the lowest 
frequencies. Unlike what is shown in the previous figures, we see that the signal is fainter at low frequencies than that at high frequencies. 
The emission becomes clearly visible in the 857 GHz channel while still present at lower frequencies. One could therefore cross-correlate all 
channels to constrain the dark matter mass.
The same feature can be seen in Fig.~\ref{fig:6muG_143GHz_Med_200GeV_ann} when one increases the magnetic field. 
However, the signal is brighter and slightly more concentrated towards the galactic centre. Again, this was to be expected since a 
large value of the magnetic field confines the electron in the galactic centre. As a result, the synchrotron emission is brighter 
but also more confined towards the centre. 

The emission is easier to observe when the propagation parameters correspond to the MAX set. In this case, it is broader (cf Fig.~\ref{fig:3muG_143GHz_MAX_200GeV_ann}).
However, in terms of intensity, it is quite similar to the MED set of parameters.

\begin{figure}
	\centering
		\includegraphics[width=0.45\textwidth]{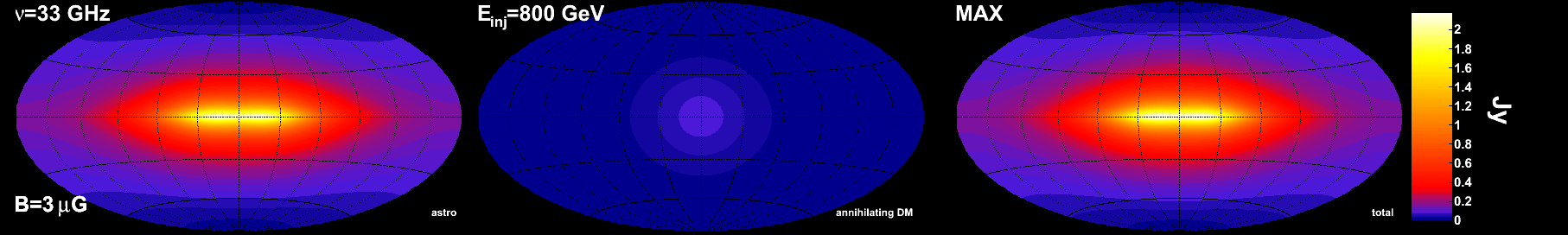}
		\includegraphics[width=0.45\textwidth]{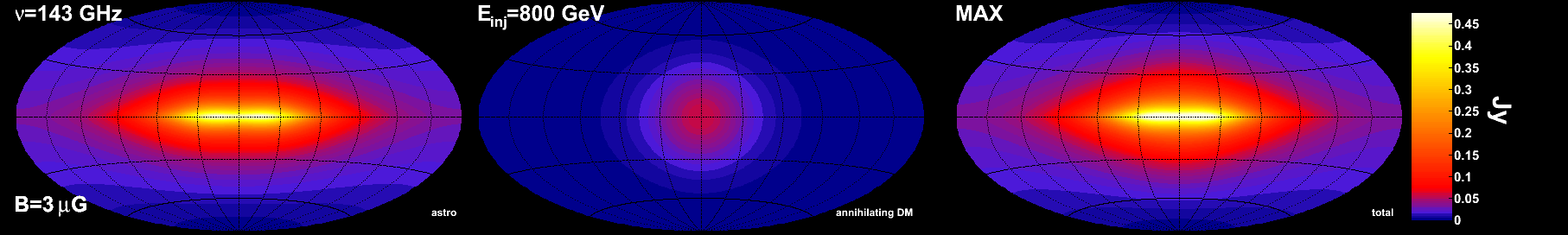}
		\includegraphics[width=0.45\textwidth]{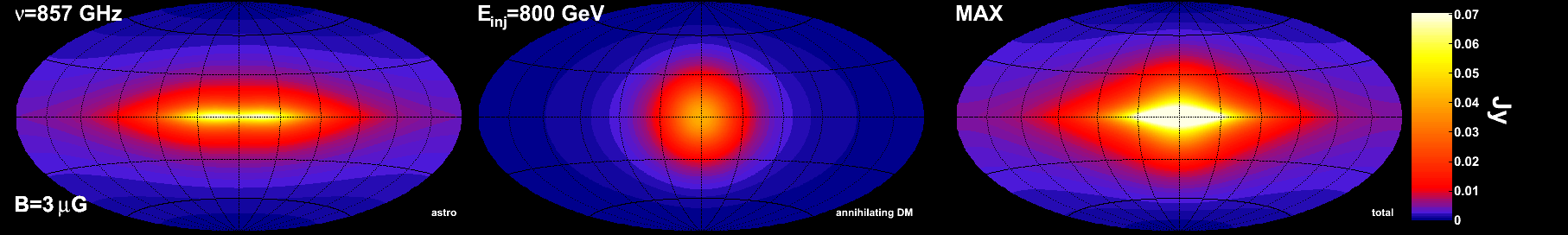}
	\caption{Synchrotron maps for 800 GeV dark matter particles, $B=3 \mu$G. We use the MAX parameter set and assume annihilating particles. }
	\label{fig:3muG_143GHz_MAX_800GeV_ann}
\end{figure}

Finally, it is interesting to note that the extinction of the dark synchrotron emission at low frequencies is particularly visible when the dark matter mass is about 800 GeV (cf Fig.\ref{fig:3muG_143GHz_MAX_800GeV_ann}). 
In this case, the LFI should not see any signal while HFI could in principle have a detection. The emission at 857 GHz should be about $7 \ 10^{-2}$ Jy. This is quite faint but the synchrotron emission associated 
with astrophysical sources is comparable. Hence, the ability for HFI to determine whether there is a ``dark'' synchrotron signal depends on the level of accuracy required to remove the other foregrounds. 
These figures demonstrate that extrapolating radio maps to high frequencies can lead to the wrong conclusions since very high energy electrons can, depending on their injection energy, shine at the highest frequencies only.

Concerning decaying dark matter, the emission is spatially much broader and because the decay rate is constrained by local cosmic-ray fluxes to be quite low (1--10 $\times 10^{-28}$ s$^{-1}$, it appears to be very difficult to distinguish from the astrophysical background. Nearby galaxy cluster observations by Fermi \citep{2010JCAP...12..015D,2011PhLB..698...44K} provide strong constraints on gamma rays from $b,\bar b$ and $\mu,\bar\mu$ channels for  decaying dark matter because of the relatively broad emission profile, and it  might be of interest to reexamine  the implications of Planck data for constraining dark matter via leptonic decays in these systems.
%===========================================================================
%===========================================================================
\section{Conclusion \label{section3}}
%===========================================================================
%===========================================================================

In this Letter, we have investigated the synchrotron emission from annihilating and decaying dark matter particles and predict the morphology 
and the intensity of the emission at PLANCK frequencies. To avoid considering unrealistic scenarios, we have required that the sum of the dark matter 
and astrophysical source contributions to the electron plus positron flux observed at Earth position be compatible with the FERMI data. 
By comparing the different LFI and HFI frequency channels, we found that the dark matter (synchrotron) signature has very specific features that could be 
used to find such a signal if it exists. For reasonable values of the magnetic field (and assuming it is uniform), we find that heavy dark matter particles 
illuminate high frequencies but do not shine in the lowest frequencies. Alternatively, light particles  ``illuminate'' the lowest frequencies (rather than the highest frequencies) 
for small values of the magnetic field. 

Although this is somewhat obvious, this characteristic indicates that the combined analysis of both LFI and HFI could help in determining the dark matter mass if an anomalous 
synchrotron emission were detected by the LFI at high latitude and absent from HFI data, or vice versa. This confirms that  the PLANK experiment has the capacity to compete with dark matter direct 
(as well as other indirect) detection experiments.  Such an analysis would be particularly useful in light of the recent claims of signals in several direct detection experiments 
\citep{2011PhRvL.106m1301A,Bernabei:2008yi,Bernabei:2010mq,2010Sci...327.1619C,2011arXiv1103.4070E}  whose (highly speculative)  interpretation as dark matter inelastic scattering events  would favour relatively light dark matter particles and could therefore be within reach of PLANCK sensitivity. 

Disentangling the dark matter signal from astrophysical sources would be rather difficult. For example a 10 GeV dark matter particle 
would contribute to both the radio and submillimeter frequency ranges. However, a combined analysis of all the different frequency channels together with improved modelling of known astrophysical sources and the propagation of cosmic rays may actually help to discriminate among various scenarios.
In any case, performing such an analysis using PLANCK data, would certainly complement the constraints on the dark matter parameter space already obtained by several dark matter experiments, including FERMI searches for $\gamma$-rays from dSph.

\section{Acknowledgment}
We would like to thank T. Ensslin for useful discussions.
This work was supported by the CNRS PICS "Propagation of low energy electrons" and 
the Spanish MICINN’s  Consolider-Ingenio 2010
Programme under grant CPAN CSD2007-00042. We also acknowledge the support of the
MICINN under grant FPA2009-08958, the Community of Madrid under grant
HEPHACOS S2009/ESP-1473, and the European Union under the Marie Curie-ITN
program PITN-GA-2009-237920.

\bibliographystyle{mn2e}
\bibliography{planck}
\end{document}